\documentclass[a4paper,11pt]{article}
\pdfoutput=1 
\usepackage{jinstpub} 
\usepackage{graphicx}
\usepackage{float}
\usepackage{caption}
\usepackage{array}
\usepackage{subfigure}
\usepackage{multirow}
\usepackage{siunitx}
\usepackage{xfrac}
\usepackage{verbatim}
\usepackage{pifont}
\usepackage{arydshln}
\usepackage{paralist}                                   
\usepackage{enumitem}
\usepackage{layout} 
\usepackage{amsmath}
\usepackage{csquotes} 
\usepackage{rotating}
\usepackage{adjustbox}
\usepackage{feynmp}
\usepackage{url}
\usepackage{hyperref}
\DeclareSIUnit\torr{Torr}
\DeclareSIUnit\atmosphere{atm}
\DeclareSIUnit\cee{c}
\DeclareSIUnit\electronvoltnr{eV_r}
\DeclareSIUnit\atomicmassunit{amu}
\DeclareSIUnit\inch{"}
\DeclareSIUnit\year{yr}
\DeclareSIUnit\townsend{Td}

\title{\boldmath Performance of 20:1 multiplexer for large area charge readouts in directional dark matter TPC detectors}
\author[1]{A. C. Ezeribe,\note{Corresponding author.}}
\author{M. Robinson,}
\author{N. Robinson,}
\author{A. Scarff,}
\author{N. J. C. Spooner}
\author{ and L. Yuriev}

\affiliation{Department of Physics and Astronomy, University of Sheffield, Sheffield, S3 7RH, U.K.}

\emailAdd{a.ezeribe@sheffield.ac.uk}

\abstract{More target mass is required in current TPC based directional dark matter detectors for improved detector sensitivity. This can be achieved by scaling up the detector volumes, but this results in the need for more analogue signal channels.  A possible solution to reducing the overall cost of the charge readout electronics is to multiplex the signal readout channels. Here, we present a multiplexer system in expanded mode based on LMH6574 chips produced by Texas Instruments, originally designed for video processing. The setup has a capability of reducing the number of readouts in such TPC detectors by a factor of 20. Results indicate that the important charge distribution asymmetry along an ionization track is retained after multiplexed signals are demultiplexed.}


\begin{document}
\maketitle
\flushbottom

\section{Introduction}\label{intro:chapterthree}
The desire to scale-up directional dark matter (DM) TPCs (time projection chambers) with low energy thresholds has been building in recent years \cite{Mayet2016, Ahlen2010, Ezeribe2017, Burns2017}. A ton-year scale directional DM detector is essential to reach the required sensitivity for DM-neutrino background discrimination beyond the so-called neutrino floor \cite{Grothaus2014,Ohare2015}. The neutrino floor is a parameter space where solar, atmospheric and diffused supernovae neutrino backgrounds are expected in direct DM search experiments \cite{Grothaus2014,Ohare2015,Ohare2016}.  Also, in a case of a DM detection claim, a directional detector would be necessary to confirm the Galactic origin and anisotropic nature of the signal \cite{Cushman2013}.  However, the cost of electronics readout required to build this massive TPC detector for directional DM detection is an issue for the technology.  For instance, the proposed CYGNUS-10 detector with a fiducial volume of 10~\si{\cubic\meter} could potentially result in about 10$^5$ channels.

\begin{figure*}[h!]
\centering
\includegraphics[clip, trim=1cm 8cm 1cm 8.2cm,width=.6\textwidth,height=0.6\textheight,keepaspectratio]{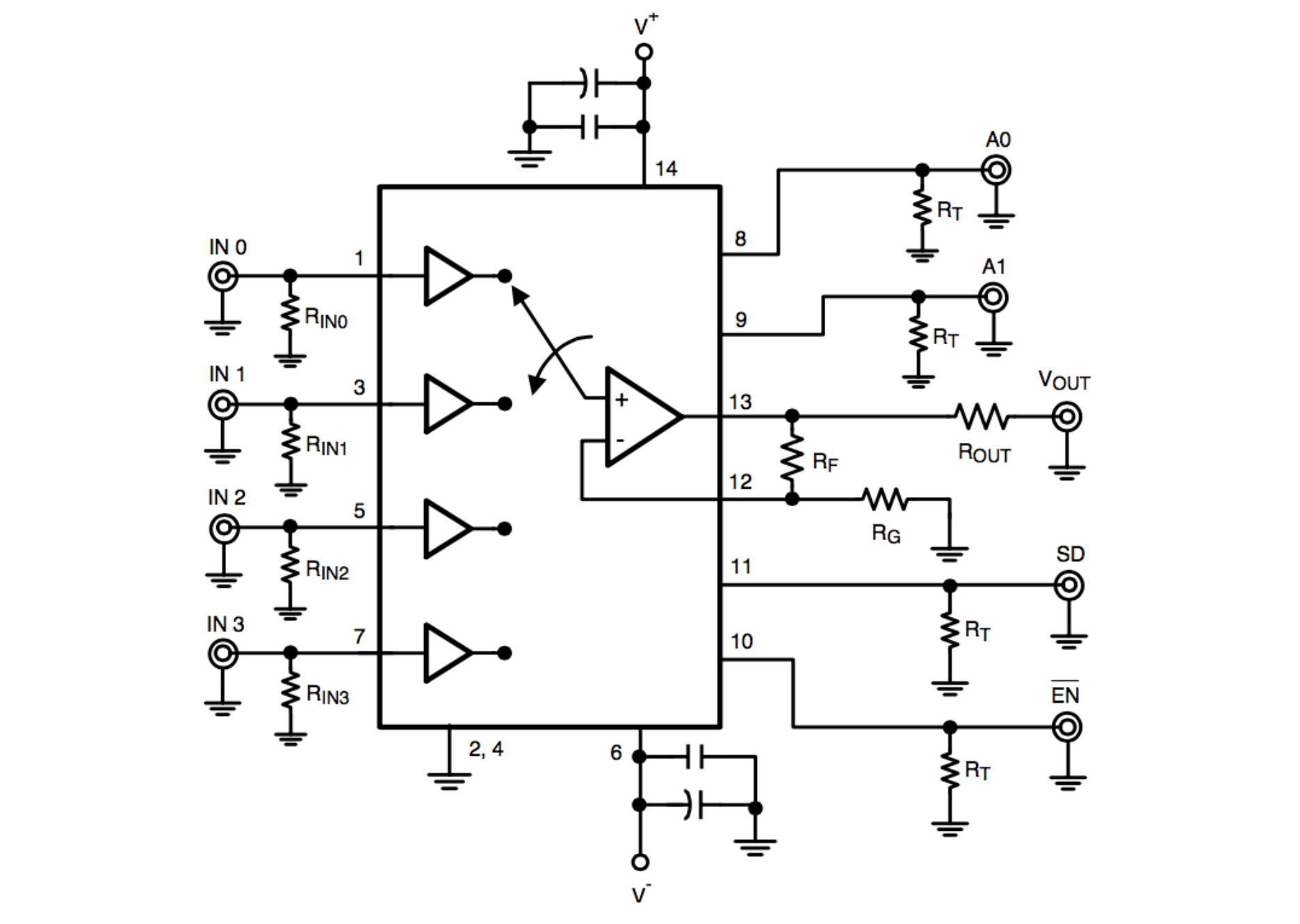}
\caption{Illustration of the 4:1 analogue multiplexer. The IN~0, IN~1, IN~2 and IN~3 are the four analogue input signal channels while A0 and A1 are the digital signals (addresses) for selecting a channel to be sampled. The switching nature of the output stream of the chip between the input channels is illustrated with an arrow \cite{LMH65742014}.}
\label{fig:muxchipprinciples}
\end{figure*}

One of the possible ways of reducing the readout cost is through signal multiplexing.  In this work, we test a 20:1 multiplexer (MUX) using TI-LMH6574 chips sourced from Texas Instruments (TI) \cite{LMH65742014} in expanded mode. Analogue multiplexers are produced mainly from field-effect transistors (FET), known as FET analogue switches which allow many signal channels to be sampled and combined into a common signal stream at a given temporal interval \cite{Horowitz2015, Stalling2007}. The TI-LM6574 chip illustrated in Figure \ref{fig:muxchipprinciples} is a high performance 4:1 analogue video multiplexer comprising a 14-pin device embedded in a small outline integrated circuit (SOIC) surface mount package. 
The four input channels of the TI-LM6574 MUX chip are marked as IN~0, IN~1, IN~2, and IN~3 in Figure \ref{fig:muxchipprinciples}.  During an operation, the signal in each of these input channels is selected using a unique signal address generated with a pair of A0 and A1 digital control signals at a defined frequency and passed as the output of the chip \cite{LMH65742014}. For the experiment described here, the frequency of the A0 signal was set to be a factor of 2 larger than the A1 signal so as to generate the required set of four digital addresses for enabling each of the analogue input channels. The generated A0,A1 digital addresses for IN~0, IN~1, IN~2 and IN~3 are 1,1~;~0,1~;~1,0~and~0,0, respectively. To recover the original signal, the multiplexed signal is demultiplexed using the reference multiplexing frequency. This was achieved by using NI 5751 digitizer adapter module \cite{ni5751-2015} of 16 ADC channels, operated with a PXI-7953R NI FlexRIO field programmable gate array (FPGA) device \cite{niflexrio} from National Instruments (NI) Corporation.

\section{Design and construction of the 20:1 signal MUX}\label{sec:multiplexer} 
\begin{figure*}[h!]
\centering
\subfigure[EAGLE schematic layout of the MUX PCB.]{
\includegraphics[clip, trim=1cm 1cm 1cm 0.05cm,width=1\textwidth,height=0.35\textheight]{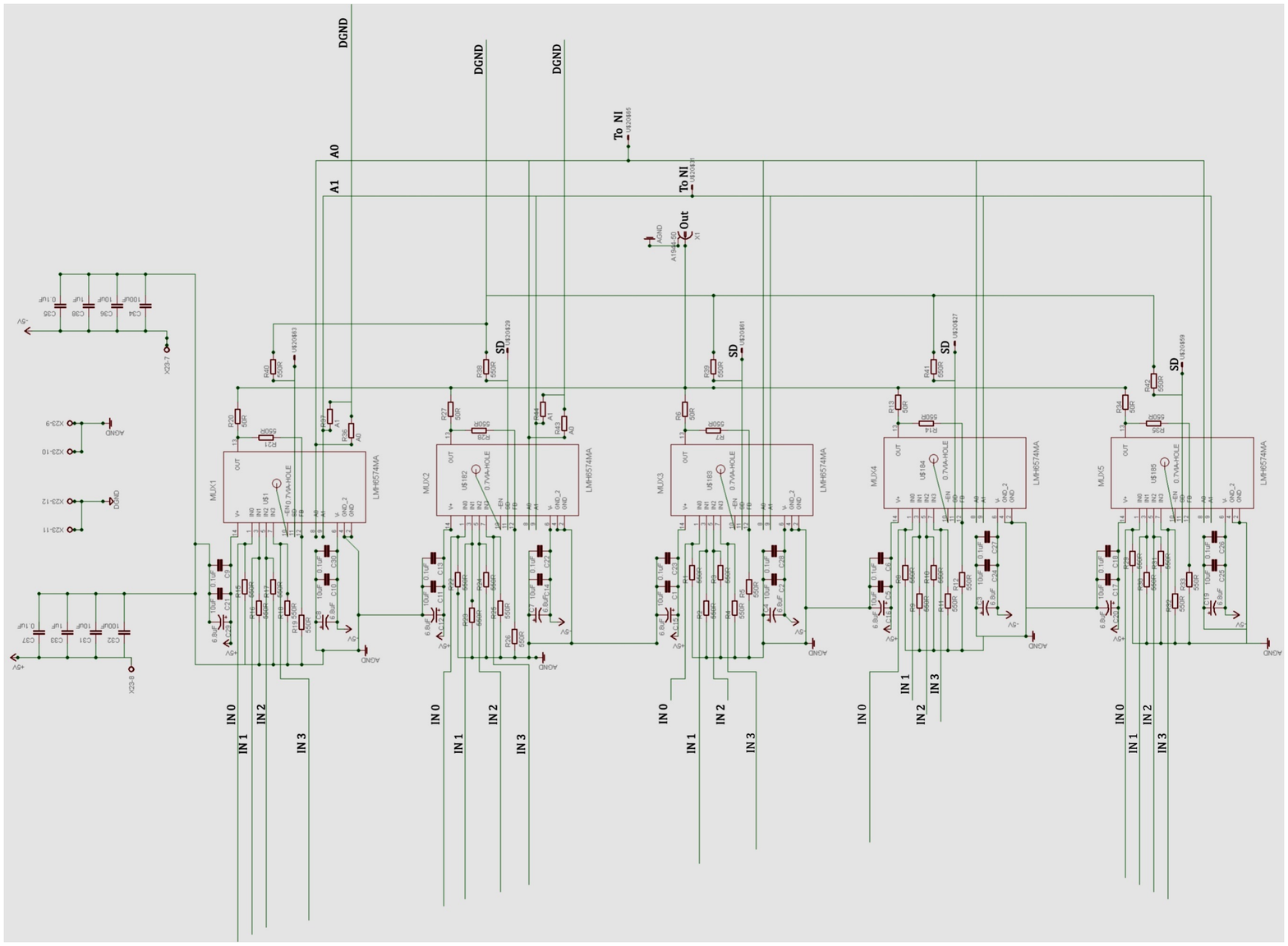} 
\label{fig:eagledesign}
}\hfil
\subfigure[Manufactured MUX PCB in a shielding box.]{
\includegraphics[clip, trim=4cm 3cm 4cm 3cm,width=0.6\textwidth,height=0.32\textheight]{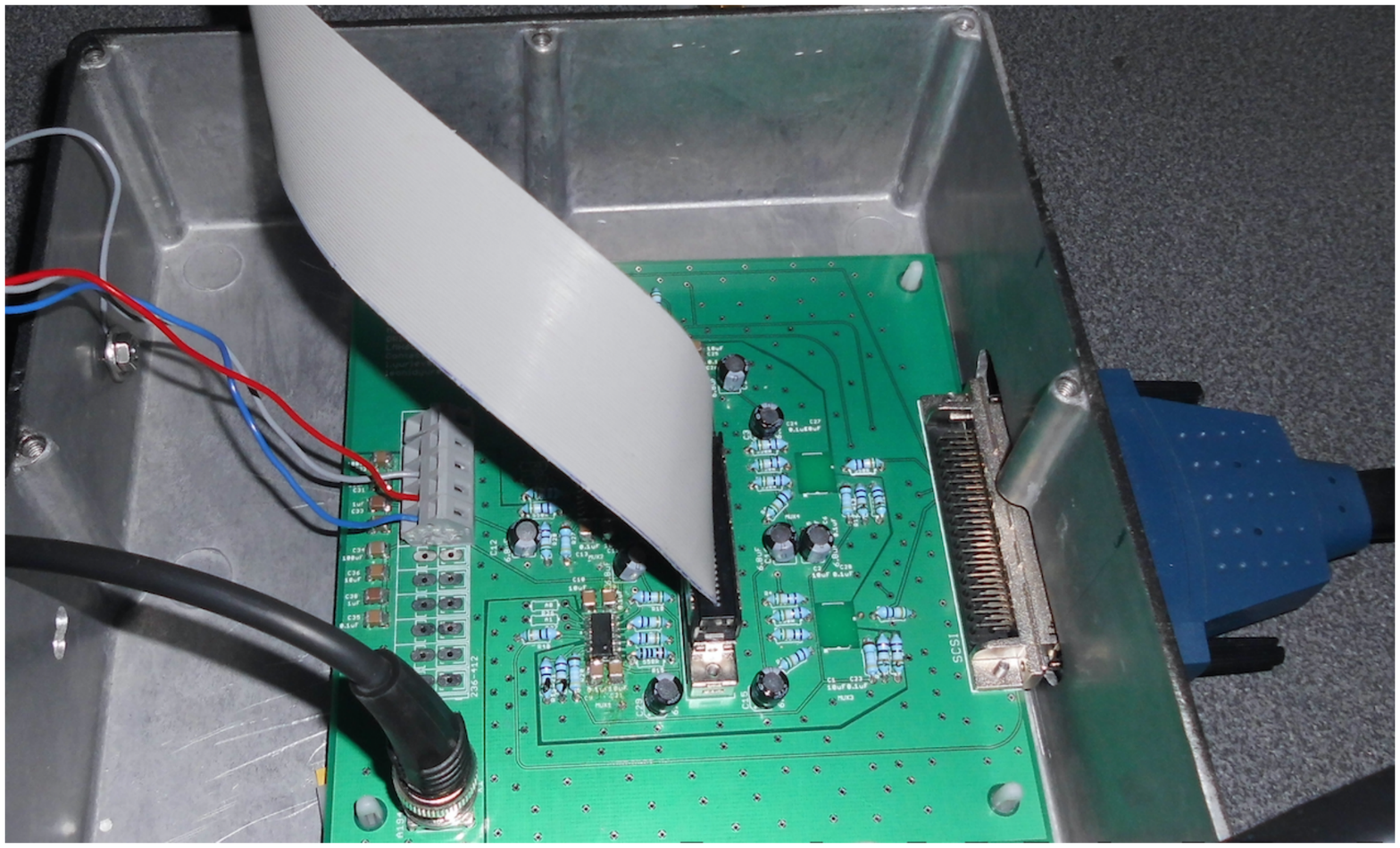}
\label{fig:manufacturedboard}
}
\caption{EAGLE schematic layout for the 20:1 analogue signal multiplexing PCB \protect\subref{fig:eagledesign} and the manufactured board in a shielding box \protect\subref{fig:manufacturedboard}.  The five light red rectangular chips in \protect\subref{fig:eagledesign} are the TI-LMH6574 chips while the resistors are shown as smaller dark red rectangles. Analogue inputs of each of the chips are labelled IN~0, IN~1, IN~2, IN~3 while the output connection is marked as Out. Digital chip control signals are marked as A0, A1 and SD. The AGND and DGND are connections to the analogue and digital grounds, respectively.}
\label{fig:muxboard}
\end{figure*}

To multiplex 20 analogue signal channels, five TI-LMH6574 chips were used. This was achieved by connecting together all the digital address (A0 and A1) pins of the chips using a custom-made EAGLE \cite{Cadsoft2014} printed circuit board (PCB). This is also true for the shutdown (SD) and the grounded chip enable ($\overline{\text{EN}}$, see Figure \ref{fig:muxchipprinciples}) signal pins to allow for centralised chip control and signal read out. This is known as expanded mode operation. The EAGLE schematic layout view of the MUX~PCB and the manufactured 20:1 MUX board are shown in Figure \ref{fig:muxboard}. 
The SD signal is the chip switching digital signal used for moving controls from one chip to another while in the expanded mode. The five light red rectangular chips in Figure \ref{fig:eagledesign} are the TI-LMH6574~MUX chips. Analogue signal inputs were connected on the board using an MDR-50 connector via a ribbon cable (see Figure \ref{fig:manufacturedboard}). The analogue signal output of the board was read out using a BNC cable. To route the chip digital control signals from the NI module to the board, a NI SCSI-68 connector and NI SHC68-68-EPM cable were used. The MUX board was powered through a set of WAGO 236-412 wire-to-board terminal connectors as shown in Figure \ref{fig:manufacturedboard}.

Each of the input signal channels was terminated with a 550~\si{\ohm} resistor to reduce signal reflections and ensure that any excess currents were properly grounded. These input termination resistors are marked as R$_{\text{IN0}}$, R$_{\text{IN1}}$, R$_{\text{IN2}}$, R$_{\text{IN3}}$, R$_\text{T}$ and R$_\text{G}$ in Figure \ref{fig:muxchipprinciples}. Also, a 550~\si{\ohm} resistor was used as the R$_\text{F}$ resistor to achieve a gain of $\sim$2, with an R$_{\text{OUT}}$ resistor of 50~\si{\ohm} on the output channel. 

\begin{figure*}[h!]
\centering
\includegraphics[clip, trim=2.4cm 7cm 4.5cm 7cm,width=1\textwidth,height=0.23\textheight]{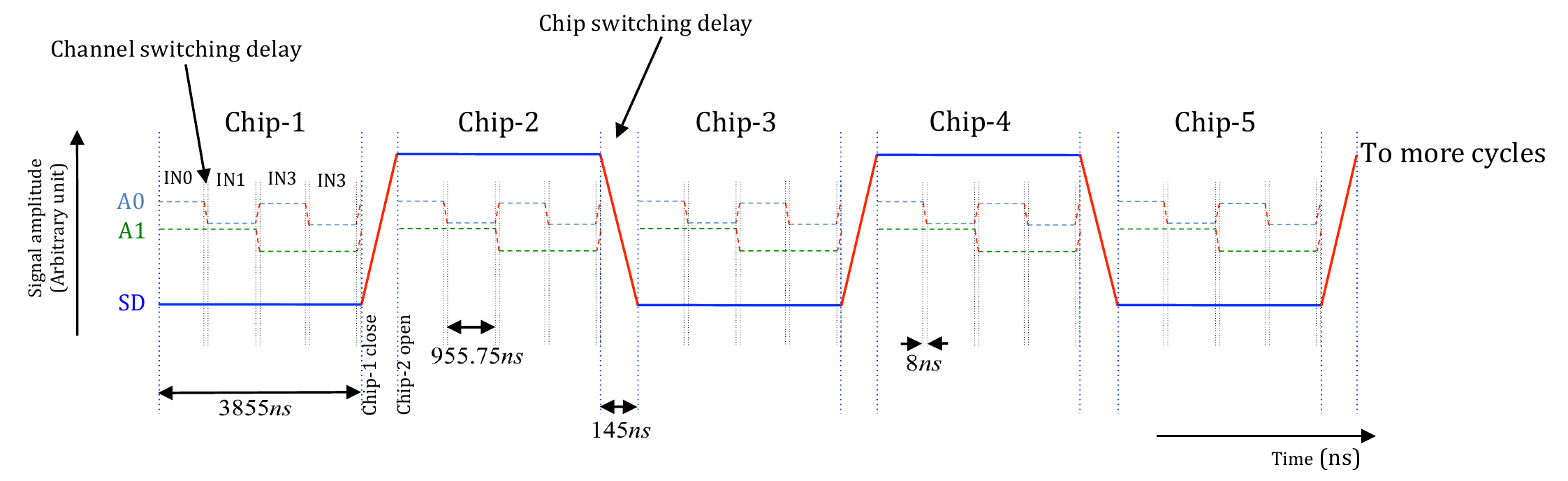} 
\caption{Temporal illustration of the 20:1 multiplexer data acquisition system (DAQ) at a sampling frequency of 1~\si{\mega\hertz} per channel. The light blue, green and blue lines are the A0, A1, and shutdown (SD) digital control signals while red lines show channel and chip switching delays. DAQ dead times due to channel and chip control switching delays are shown with vertical black and blue doted lines, respectively.}
\label{fig:timingdiagram}
\end{figure*}

To achieve the required operational specification, the analogue ground was separated from the digital ground to avoid noise coupling between the analogue and digital signals. However, when this is not properly managed it can introduce crosstalk in the circuit since both planes may radiate noise or act as noise antennas.  Such noise is due to return currents flowing beneath each signal line. This return current will always prefer to follow the path with lowest impedance hence disconnecting these return paths with separated grounds may result in potential current loops especially when there are signal traces over the ground breaks.  A circuit with a resultant current loop can experience high ground inductance which is susceptible to signal interference.  In this design, the issue of crosstalk induced by separated analogue-digital grounds was avoided by not running traces across the two grounds. The ground-signal-ground trace stack-up arrangement was adopted to reduce electromagnetic interference (EMI) on the board.  Micro-strip trace technology was used in this design due to its characteristically lower dielectric losses, easy accessibility for maintenance and their low manufacturing cost relative to strip-line technology. While operating the MUX in the expanded mode, an inter-chip switching delay of about 145~\si{\nano\second} would be expected \cite{LMH65742014,Battat2015talk}. This is illustrated in Figure \ref{fig:timingdiagram}. 
The inter-chip switching delays are one of the major challenges of modular analogue multiplexers. For this MUX system, a total of about 0.89~\si{\micro\second} delay and data acquisition system (DAQ) dead time are expected considering the 8~\si{\nano\second} channel switch time of the individual chips. This result in <5\si{\percent} loss for signals of 20~\si{\micro\second} duration from Cremat CR-200-4\si{\micro\second} \cite{Cremat2014shaper} shaping amplifier. 

\section{Test system for the 20:1 MUX}\label{sec:multiplexeroperation}
\begin{figure*}[b!]
\centering
\subfigure[Schematics of the TPC detector.]{
\includegraphics[clip, trim=2cm 8cm 2cm 5cm,width=0.41\textwidth,height=0.27\textheight]{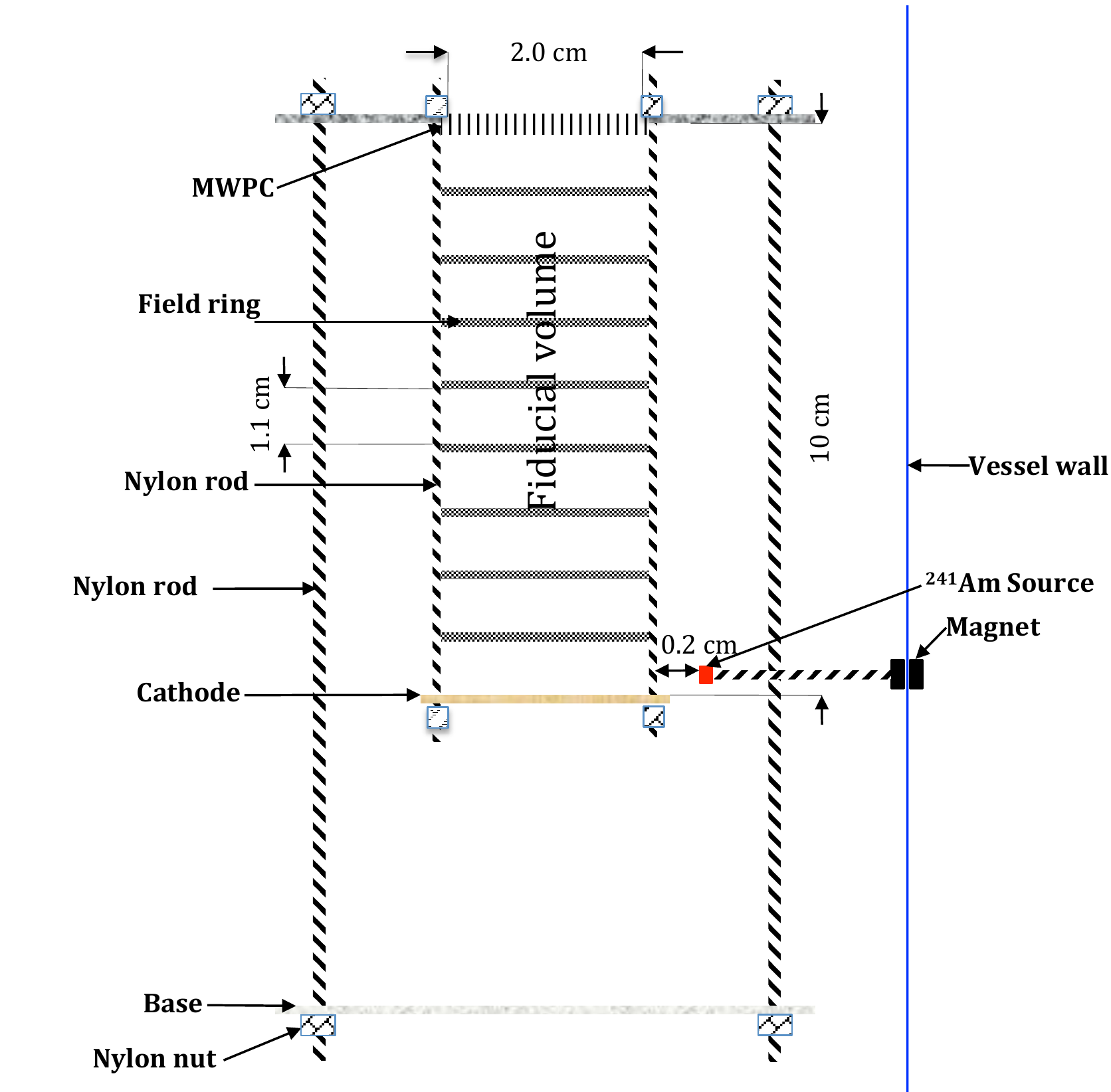}
\label{fig:detectorshemematics}
}\hfil
\subfigure[Constructed TPC detector.]{
\includegraphics[clip, trim=1cm 5cm 1cm 3cm,width=0.31\textwidth,height=0.27\textheight]{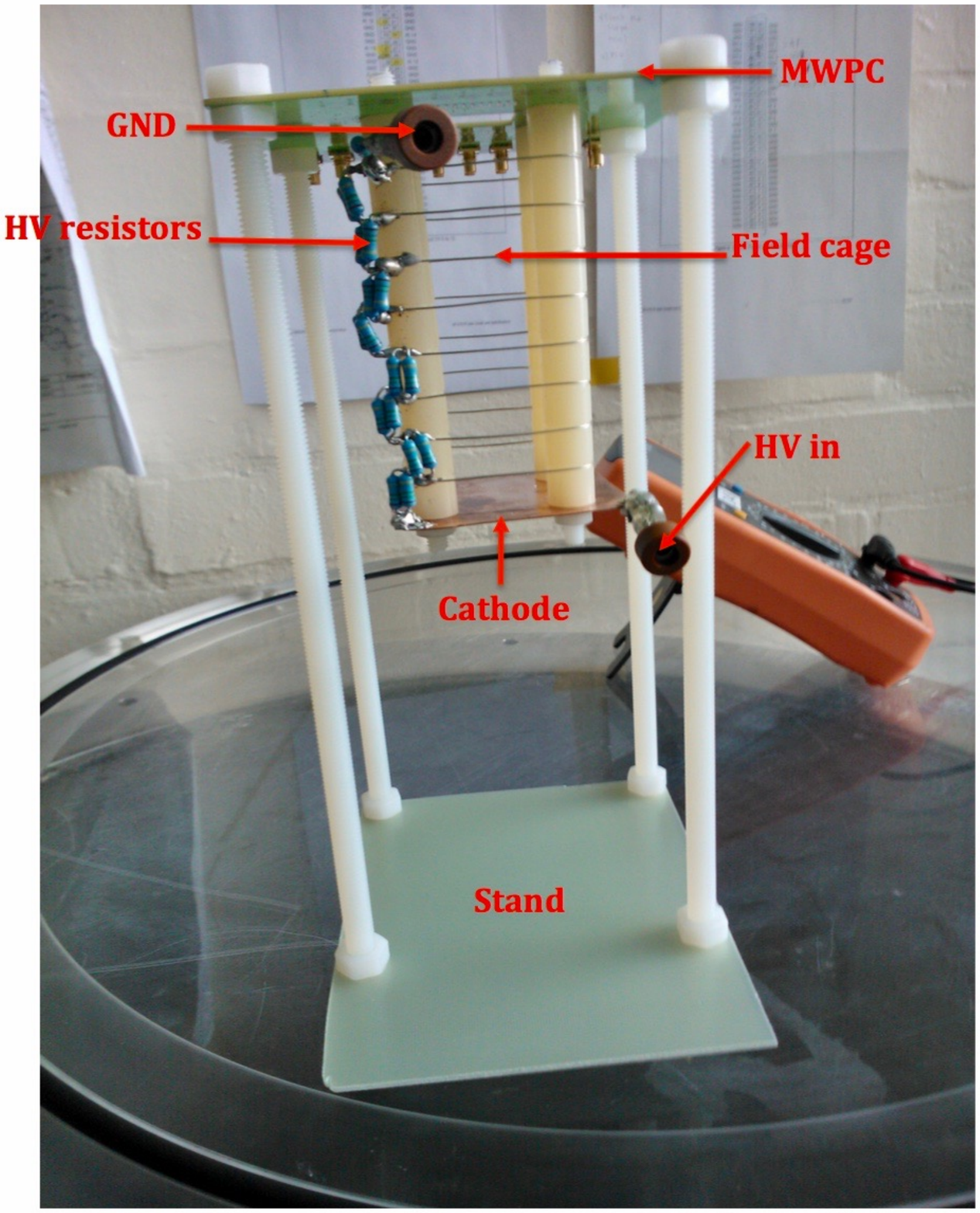}
\label{fig:manufactureddetector}
}\hfil \hfil \hfil
\hfil
\subfigure[Circuit diagram of the one-plane MWPC readout.]{
\includegraphics[clip, trim=3.5cm 16cm 4cm 9cm,width=0.46\textwidth,height=0.11\textheight]{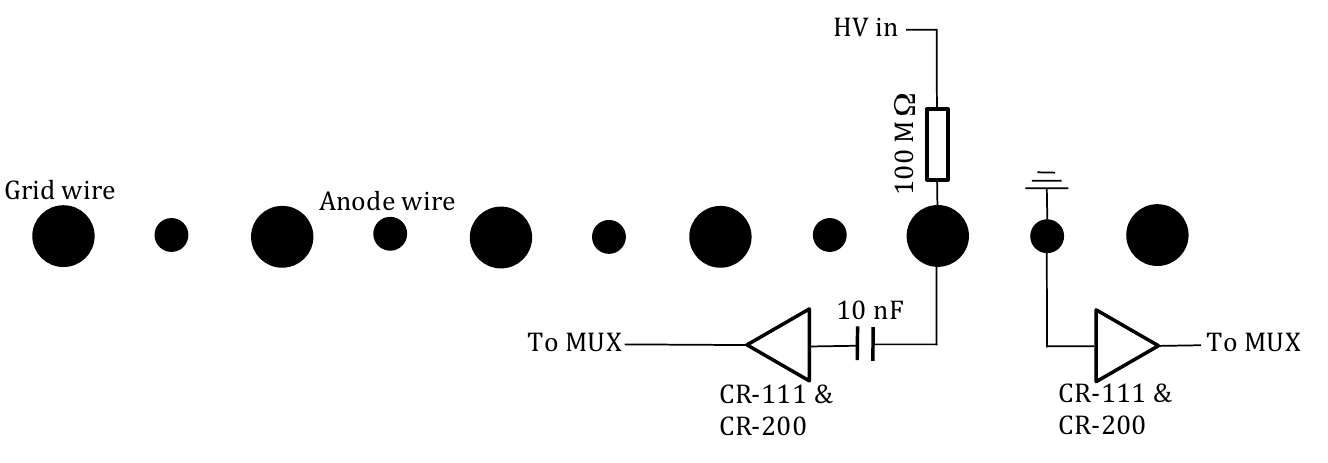}
\label{fig:mwpccirccit}
}
\caption{Miniature TPC detector used to generate analogue input signals used for testing the 20:1 MUX system. In \protect\subref{fig:detectorshemematics} is the schematics of the detector showing the position of the source,  \protect\subref{fig:manufactureddetector} is the detector after construction while \protect\subref{fig:mwpccirccit} shows the wire configuration and circuit of the detector readout.}
\label{fig:tpcdetector}
\end{figure*}

A new miniature TPC detector (shown in Figure \ref{fig:tpcdetector}) was built to test the MUX electronics using analogue signals from 5.5~\si{\mega\electronvolt} alpha interactions. 
The TPC is a 2~\si{\centi\meter}$~\times$~2~\si{\centi\meter}$~\times~$10~\si{\centi\meter} detector with a one-plane multiwire proportional chamber (MWPC) \cite{Sauli1977,Charpak1979} readout. In this one-plane MWPC design, 10 grounded stainless steel anode wires of 20~\si{\micro\meter} diameter were sandwiched between parallel 100~\si{\micro\meter} grid wires with an anode-grid pitch of 1~\si{\milli\meter}.  The alternating anode and grid wires were mounted on a customized EAGLE PCB using an automated quad wire-winding robot (QWWR) \cite{Ezeribe2014memo} to achieve uniform wire spacing. To operate the QWWR, the two wire spools were attached to two stepper motors. Wire from each of the spools was passed through a 1~\si{\milli\meter} pitched double grooved pulley. Wires from the pulley were then tensioned and fixed on one of the four 1~\si{\milli\meter} threaded wire-landing rods positioned on a wire-winding frame. Two ball bearings were used to attach the remaining opposite sides of the cubic winding frame to a U-shaped holder.  After the first 360\si{\degree} turn of the winding frame, the ball bearing aided mobile wire tensioning and unspooling system driven by a C++ based DAQ was set to undergo a 2~\si{\milli\meter} displacement to repeat the process. The accuracy (precision) of the QWWR is $\pm$0.002~\si{\milli\meter} ($\pm$0.015~\si{\milli\meter}) as determined from data obtained with a BS-6020TRF microscope.  This result is consistent with the accepted range in Refs. \cite{Chiba1980,Kuze1987} for optimal performance in MWPCs. During the wire-winding process, the anode and  grid wires were tensioned with 20~\si{\gram} and 130~\si{\gram} masses, respectively. This is to achieve a uniform linear density of $\sim 10^{6}$~\si{\gram\per\meter} on the anode and the grid wires. 

In operation, a set high negative potential applied on each of the grid wires through a 100~\si{\mega\ohm} power resistor was used to create electron avalanche multiplication \cite{Anderson1992} on the anode wires. Signals arising from cations created in these avalanches are decoupled using 10~\si{\nano\farad} capacitors connected to the grid wires. The decoupling capacitors serve as active vetoes against unwanted DC signals and prevent the amplifiers from discharging the grid signal wires. As illustrated in Figure \ref{fig:mwpccirccit}, each of the 10 anode and 10 of their corresponding grid channels were read out through an AC coupled Cremat CR-111 \cite{Cremat2014preamp} pre-amplifier and CR-200-4\si{\micro\second} shaping amplifier, providing the required 20 analogue signal inputs to the MUX system.  The sensitive aperture of the one-plane MWPC was surrounded by 4~\si{\milli\meter} thick copper field ring biased at negative potential to minimise signal loss.  A centralised one-plane MWPC can potentially be used to read out two back-to-back TPC detectors in modular detector designs, thereby reducing the overall readout channels relative to conventional MWPC designs \cite{Sauli1977,Charpak1979}, as used in the DRIFT-IId detector \cite{Battat2017}.

\begin{figure*}[b!]
\centering
\includegraphics[clip, trim=1cm 7cm 1cm 8cm,width=.6\textwidth,height=0.6\textheight,keepaspectratio]{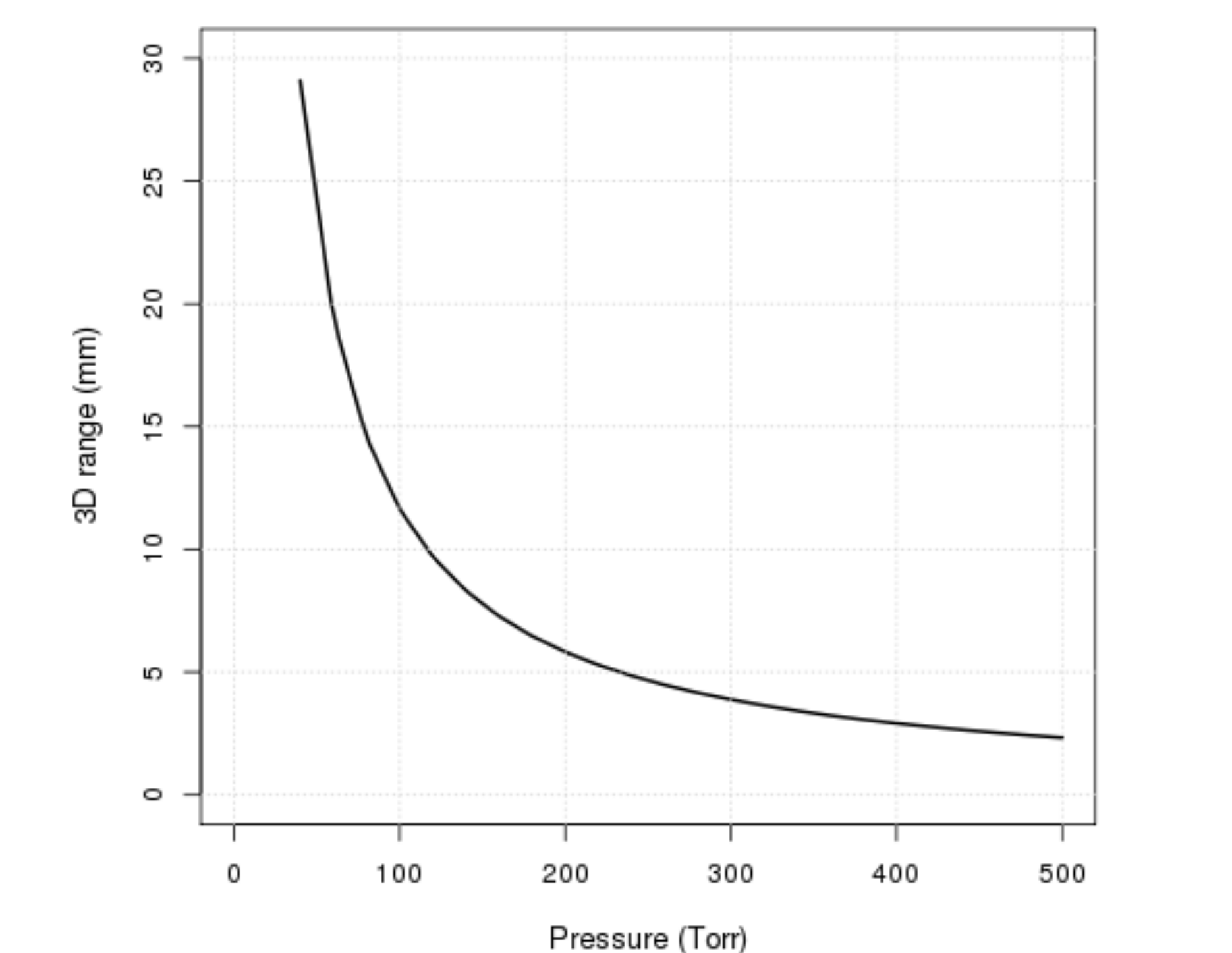}
\caption{3D range of 5.5 \si{\mega\electronvolt} alphas in various pressures of CF$_4$ gas determined from SRIM.}
\label{fig:srimresults}
\end{figure*}

The copper cathode plate was biased at -3.5~\si{\kilo\volt} to create a uniform drift field of 354~\si{\volt\per\centi\meter}. This cathode voltage was systematically stepped down through a set of eight field rings made with 0.6~\si{\milli\meter} diameter copper wires at 1.1~\si{\centi\meter} pitch and connected via a series of 750~\si{\kilo\ohm} resistors. This field ring pitch was modelled and optimised in Garfield \cite{Veenhof1984} to ensure no field leakage within the fiducial volume of the detector.  The miniature TPC detector was mounted in a 96 litre stainless steel vacuum vessel and filled with 250~\si{\torr} of CF$_4$ gas at room temperature. The expected alpha range generated from SRIM is shown as a function of CF$_4$ pressure in Figure \ref{fig:srimresults}.
\begin{figure*}[h!]
\centering
\subfigure[Illustration of the experimental set-up.]{
\includegraphics[clip, trim=7cm 6.9cm 8cm 4cm,width=0.7\textwidth,height=0.35\textheight,keepaspectratio]{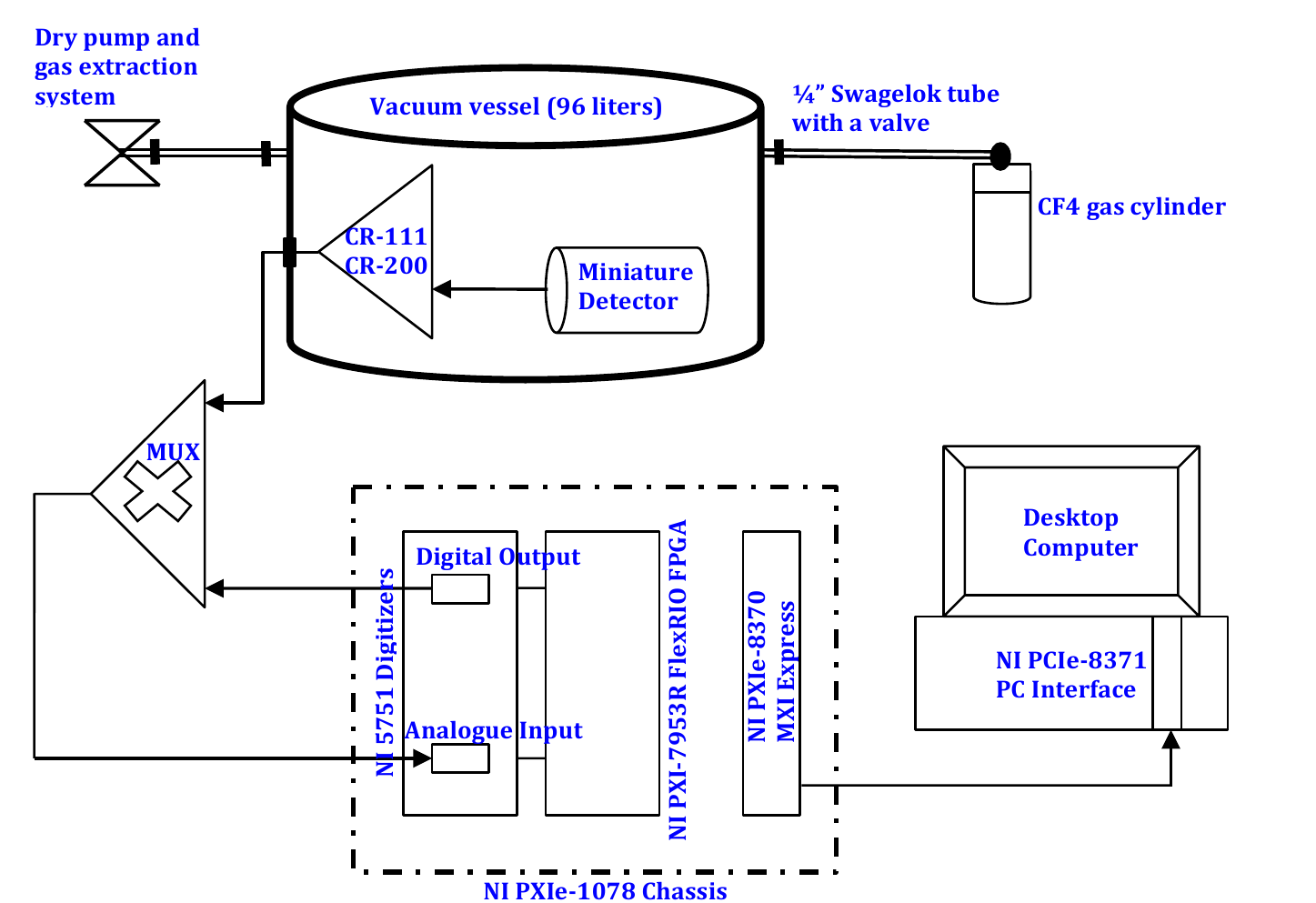}
\label{fig:diagramexperimentalsetup}
}\hfil
\subfigure[Picture of the vacuum set-up.]{
\includegraphics[clip, trim=0.5cm 3cm 0.1cm 3cm,width=0.55\textwidth,height=0.2\textheight]{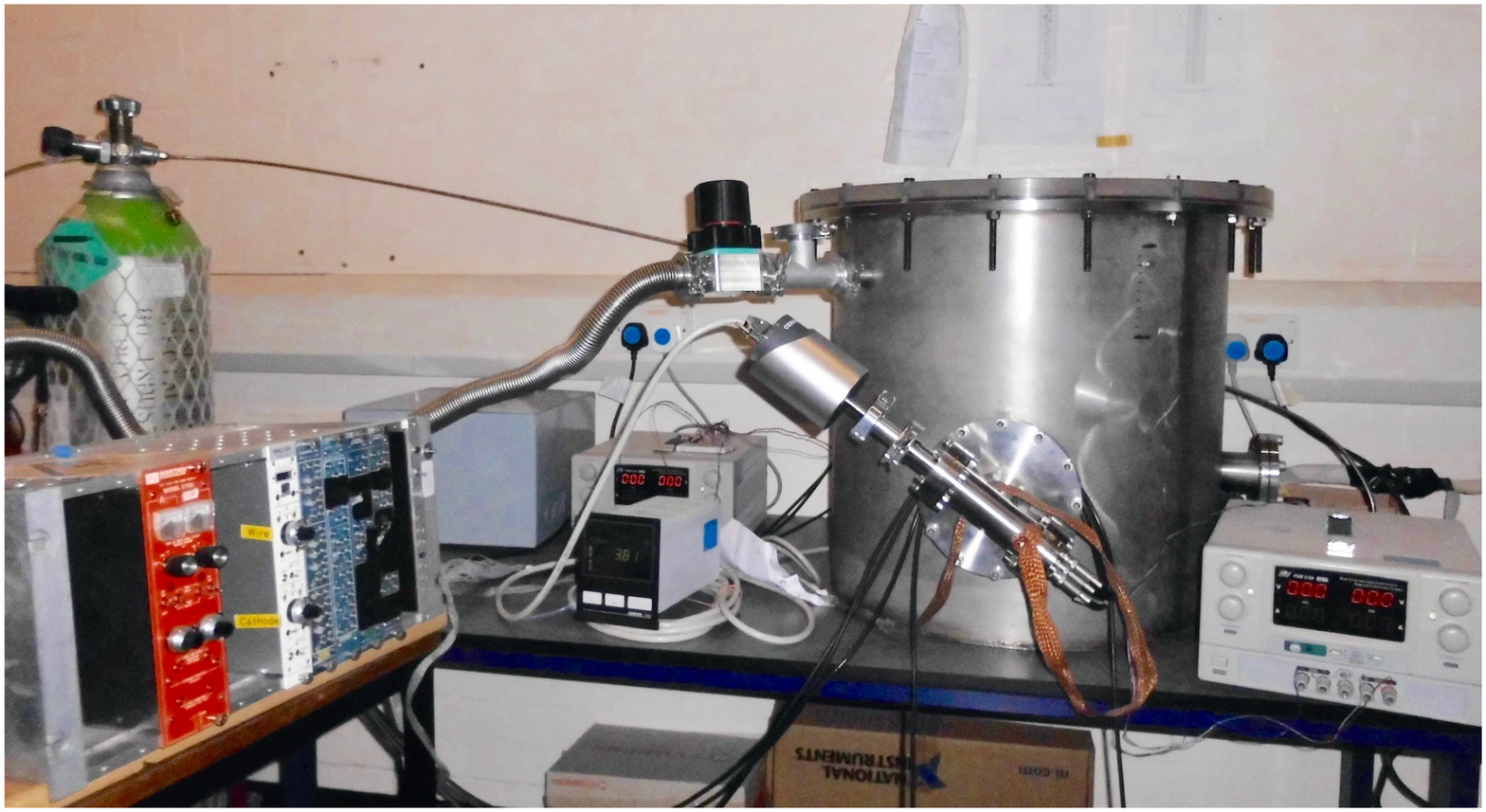}
\label{fig:vessel}
}
\caption{Experimental set-up used to test the MUX electronics. The top panel shows data flow between the major components of the set-up and gas piping while the bottom panel shows a picture of the vacuum set-up as in the laboratory. Single headed arrows in \protect\subref{fig:diagramexperimentalsetup} show directions of data flow.}
\label{fig:experimentalsetup}
\end{figure*}
It is expected that the ion drift velocity at these operational $\text{E}/\text{P}$ (1.35~\si{\kilo\volt\per\centi\meter\per\atmosphere}) and $\text{E}/\text{N}$ (5.5~\si{\townsend}) values of the detector is 10~\si{\centi\meter\per\micro\second} \cite{Vavra1993}. Here E, P and N represent the operational drift field, gas pressure and number density, respectively. The effect of gas ageing due to a contamination rate of $<$0.1\% per day of the setup should be minimal for these measurements.  The CF$_4$ gas is a common target gas used in leading directional dark matter experiments for spin-dependent sensitivity \cite{Lewin1996,Nakamura2015,Battat2016,Battat2017b}. The operational pressure of the set-up was optimised in SRIM \cite{Ziegler2010} to contain the expected Bragg peak of the alpha tracks within the fiducial volume of the detector. 

As expected, it can be seen that the range of the alpha tracks fall exponentially at higher CF$_4$ gas pressure.  Alpha tracks from a 5.5~\si{\mega\electronvolt} $^{241}$Am source were used to calibrate the detector gain, found to be 385 for 250~\si{\torr} of CF$_4$ gas. To measure this, the source was mounted on the tip of M$6$ nylon rod of 5~\si{\centi\meter} length attached to the inside wall of the vacuum vessel using a neodymium disc magnet.  Hence, by using another neodymium disc magnet on the outside wall of the vessel, the position of the source could be controlled. SRIM was used to determine the fraction of the alpha energy deposited in the detector fiducial volume. 
Before starting a new operation, the vessel was evacuated using an Edwards XDS10 scroll vacuum pump for at least 1 day.  The vessel pressure was constantly monitored using a CERAVAC CTR-101 pressure gauge acquired from Oerlikon Leybold vacuum company. A picture and illustrative diagram of the experimental set-up are shown in Figure \ref{fig:experimentalsetup}. High voltage cables used to power the cathode, field ring and the grid wires were fed into the stainless steel vacuum vessel through safe high voltage (SHV) connector feedthroughs with a maximum tolerance of 5~\si{\ampere} at a potential of 5~\si{\kilo\volt}. A PCB based feedthrough was constructed and mounted on a flange for analogue signals and amplifier power connections. 

\begin{figure*}[h!]
\centering
\subfigure[Oscillogram of one anode signal channel.]{
\includegraphics[clip, trim=2cm 7cm 2cm 9cm,width=0.47\textwidth,height=1\textheight,keepaspectratio]{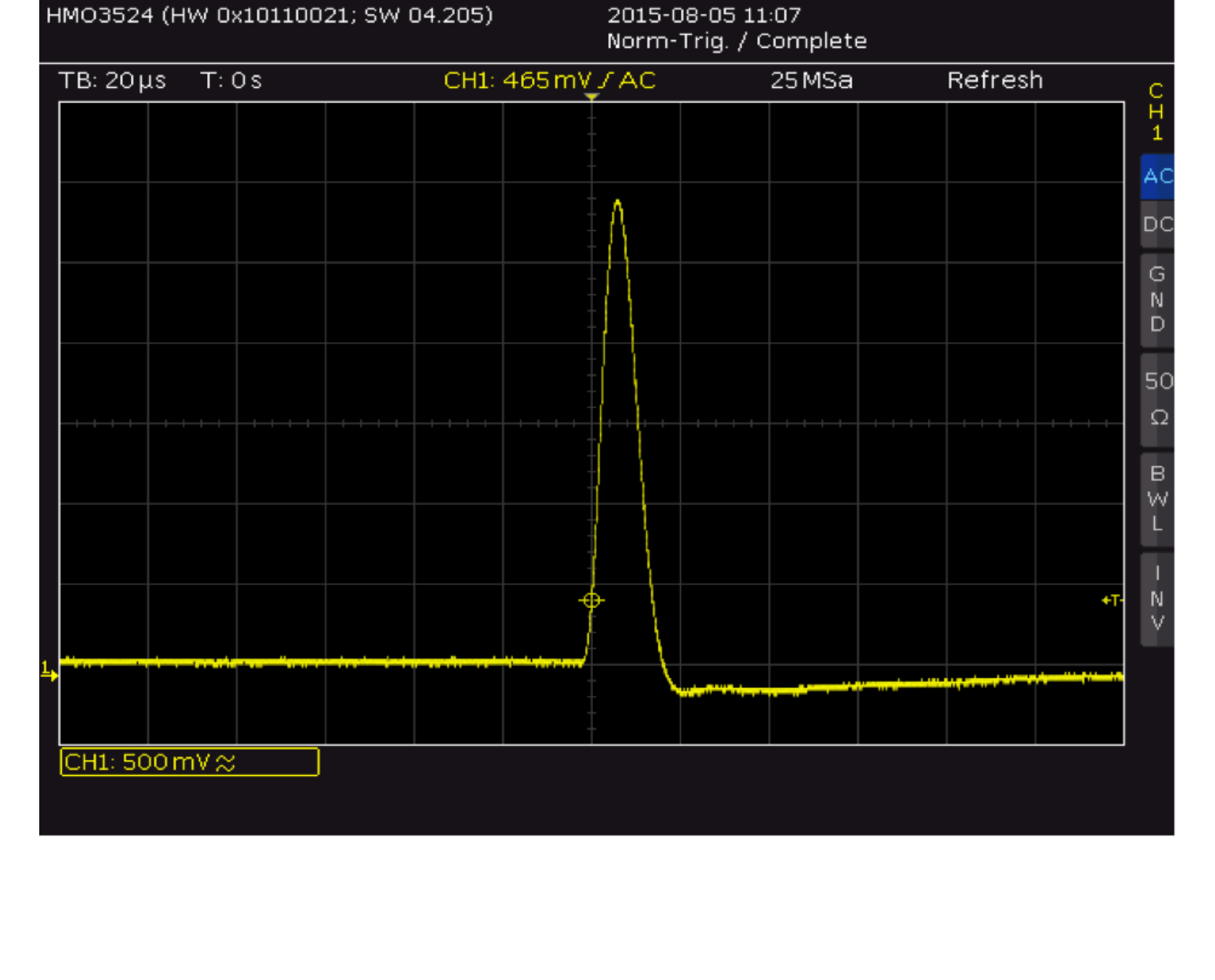}
\label{fig:scopeevent}
}\hfil
\subfigure[Detector response with varying grid potential.]{
\includegraphics[clip, trim=2cm 6.5cm 2cm 8.3cm,width=0.5\textwidth,height=0.26\textheight]{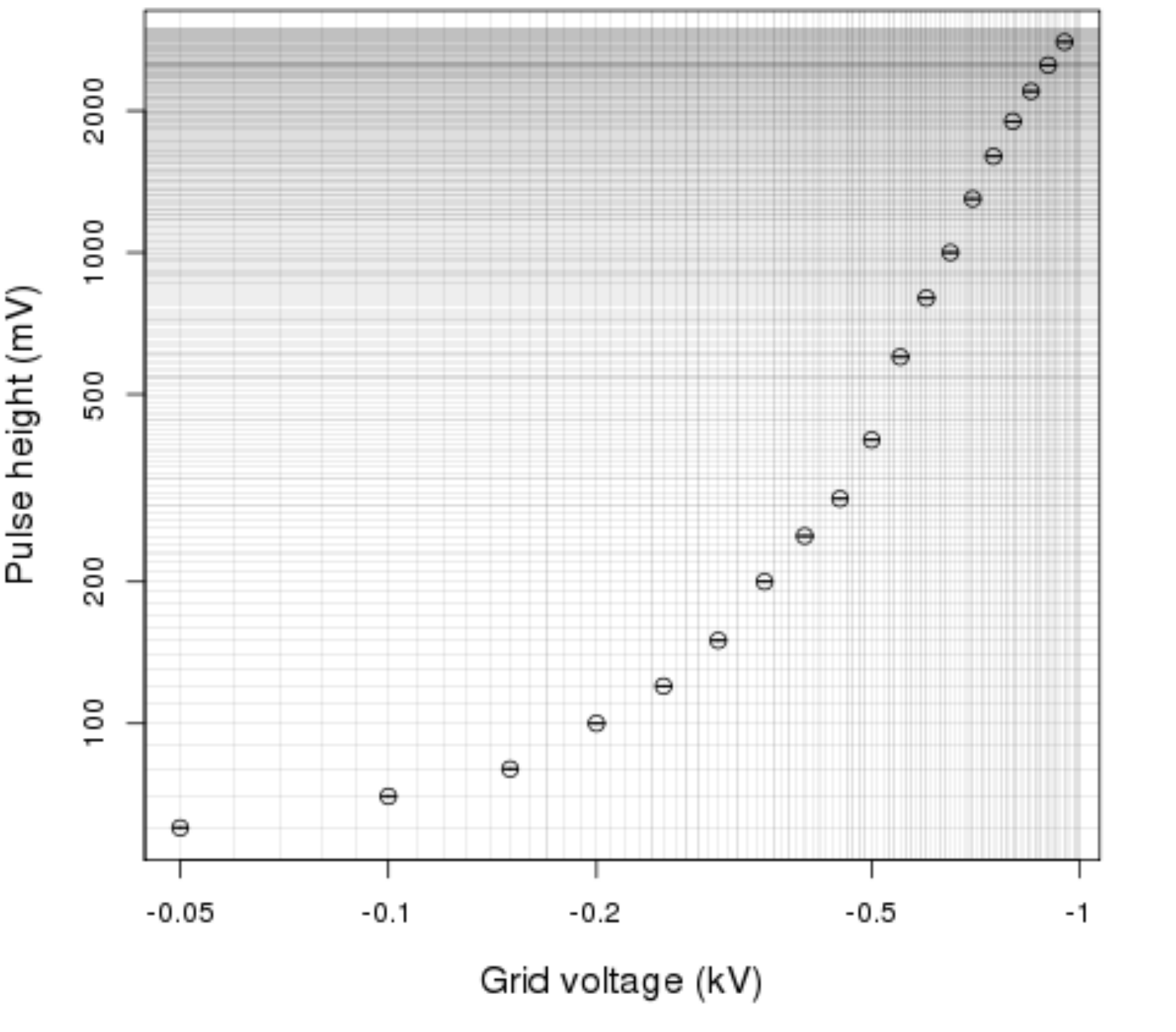}
\label{fig:detectorgain}
}
\caption{Detector response when exposed to alphas from $^{241}$Am source. In the left panel is a sample of signal observed on an anode wire while in the right panel is recorded anode pulse heights shown as a function of grid potential. The average uncertainty on the event pulse heights in \protect\subref{fig:detectorgain} is $\pm$0.2~\si{\milli\volt}. Note that the polarity of the anode channels was set to be inverted using the shaping electronics. The signal amplitude and time base of \protect\subref{fig:scopeevent} are 2.9~\si{\volt} and 20~\si{\micro\second}, respectively.}
\label{fig:detectorresponse}
\end{figure*}

An example of a typical charge signal pulse obtained from the 8$^{\text{th}}$ anode wire with the source in place is shown in Figure \ref{fig:scopeevent} when the detector is exposed to the alpha source . The polarity of this signal pulse was inverted using the polarity switch on the Cremat CR-160-R7 board \cite{Cremat2014shaperboard}. This is to ensure that the signals on the grid and anode signal wires have common polarity. The pulse height of the signal shown in Figure \ref{fig:scopeevent} is about 2.8~\si{\volt} with a resolution of $\sim$10~\si{\micro\second} and an undershoot of  $\sim$0.15~\si{\volt}, below the baseline. The magnitude of the signal undershoot observed from each alpha event depends on the fraction of the energy that is deposited on the wire under consideration.  This undershoot effect was accounted for in data analyses. 
For an overview of the anode channel response as a function of voltage of the grid wires, see Figure \ref{fig:detectorgain}. It can be seen that the maximum gains were obtained at higher (negative) grid wire voltages >$0.8$~\si{\kilo\volt}. The grid wires were set to $0.85$~\si{\kilo\volt} during these measurements to minimise the risk of wire breakage during the avalanche process. Also, signals with more than $3$~\si{\volt} amplitudes saturate the amplifier leading to signal losses.  As expected, the signal pulse heights show an exponential increase with the grid voltage especially at voltages <$0.7$~\si{\kilo\volt}. The observed increase is more apparent for grid voltages  >$0.3$~\si{\kilo\volt}. The expectation is that the signal pulse heights and gas gain will decrease at higher gas pressures. However, this is not a problem for this experiment as all the measurements were performed using constrained gas pressures (250 - 260~\si{\torr}).  

Data from the miniature detector was multiplexed and demultiplexed using an NI FPGA based LabVIEW DAQ. For details of the NI FPGA and digitizer module used in design of this DAQ, see Figure \ref{fig:diagramexperimentalsetup}.  Demultiplexed data were saved on disk in a desktop computer for analyses.

\section{MUX Results and Discussion}\label{sec:results}
\begin{figure*}[h!]
\centering
\includegraphics[clip, trim=2cm 7cm 2cm 8cm,width=0.8\textwidth,height=0.9\textheight,keepaspectratio]{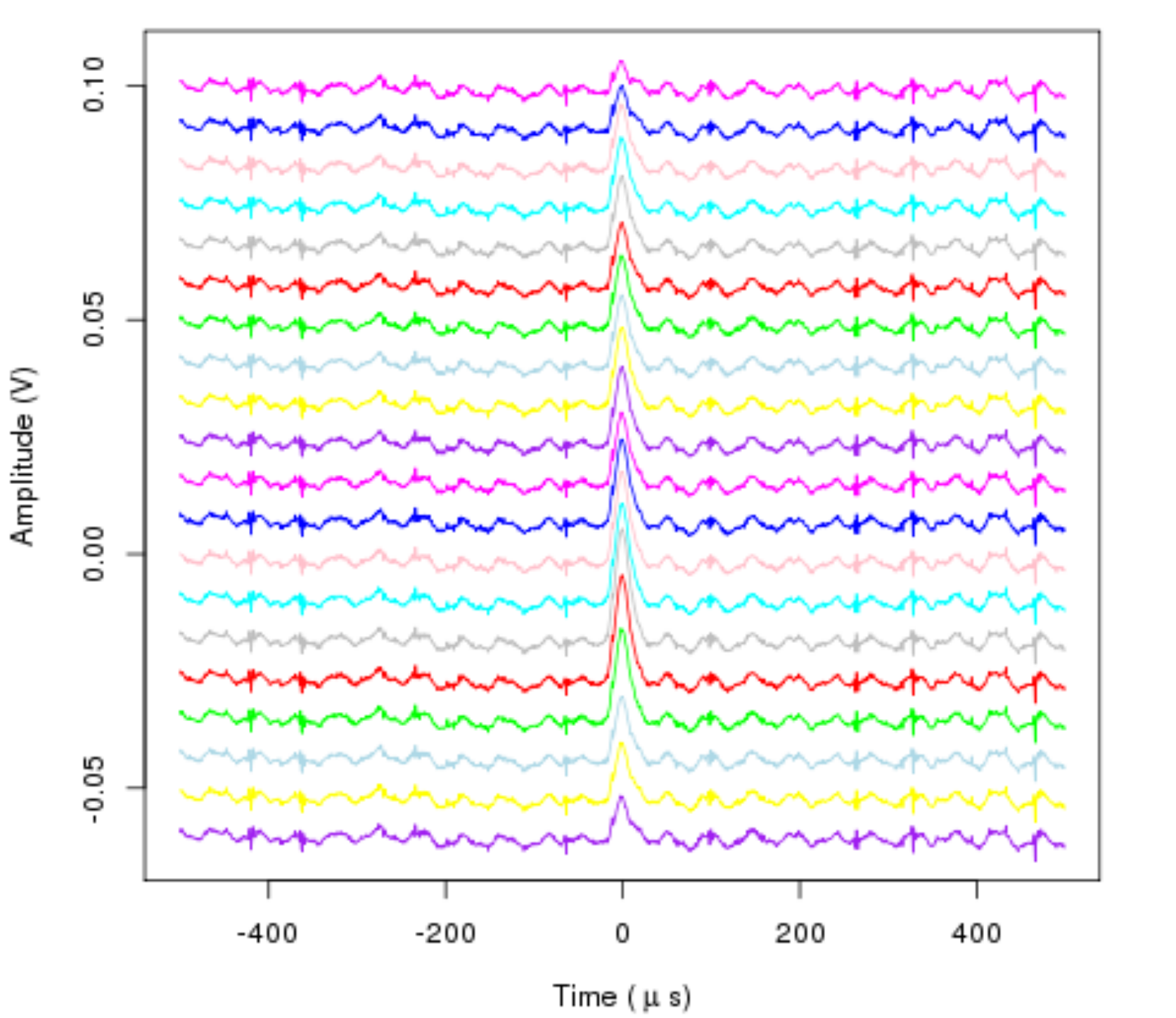}
\caption{Sample of demultiplexed analogue signals showing the response of the miniature detector when exposed to an alpha from $^{241}$Am source in 250~\si{\torr} of CF$_4$. Top side of the panel is closer to the source. }
\label{fig:demuxall}
\end{figure*}

The 20 analogue signal inputs from the one-plane MWPC based detector were multiplexed and demultiplexed at 0.625~\si{\mega\hertz} per signal channel (12.5 ~\si{\mega\hertz} for 20 channels) using one ADC channel of the NI module. A sample of reconstructed demultiplexed alpha signal track pulses is shown in Figure \ref{fig:demuxall}. 
To ensure that the Bragg peak of the alpha tracks were contained within the detector fiducial volume, 250~\si{\torr} of CF$_4$ gas was used. Signal glitches and background low frequency sinusoidal waveform noise can be seen on each of the signal channels. These glitches and low frequency ($\sim$20~\si{\kilo\hertz}) noise were not present before the analogue signals were multiplexed.  This shows that they were coupled to the signals during the multiplexing or demultiplexing processes or both. This type of noise can be suppressed in future design by isolating the power channel, analogue and digital channel grounds using different PCB ground layers to minimize the effect of noise from potential grounding loops. The quality of the signal input to the multiplexer can be improved by using coaxial cables to route signals from detector to the multiplexer board instead of the ribbon cable used in this test.  Also, shorter conducting traces can reduce the MUX board's susceptibility to radiated and other extraneous low frequency noise.  Further investigations show that the observed signal-to-noise ratio (SNR) decays with increase in the multiplexing frequency.  An exponential decay in SNR was observed when operating with multiplexing frequencies that are $>$1~\si{\mega\hertz}.

\begin{figure}[h!] 
\centering
\subfigure[Demultiplexed 686.3~\si{\milli\volt} pulse.]{ 
\includegraphics[width=0.485\linewidth,height=0.306\textheight]{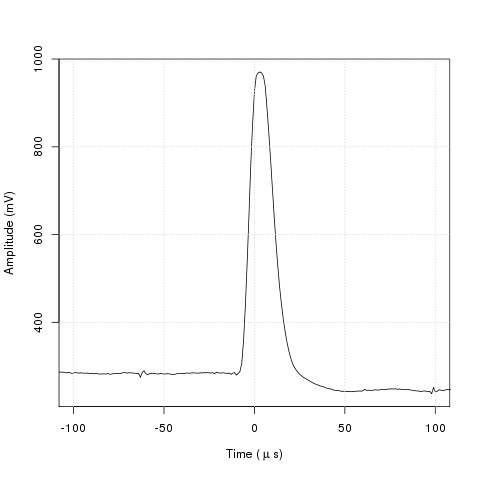}
\label{fig:demux4}
}\hfil
\subfigure[Demultiplexed 579.6~\si{\milli\volt} pulse.]{
\includegraphics[width=0.485\linewidth,height=0.306\textheight]{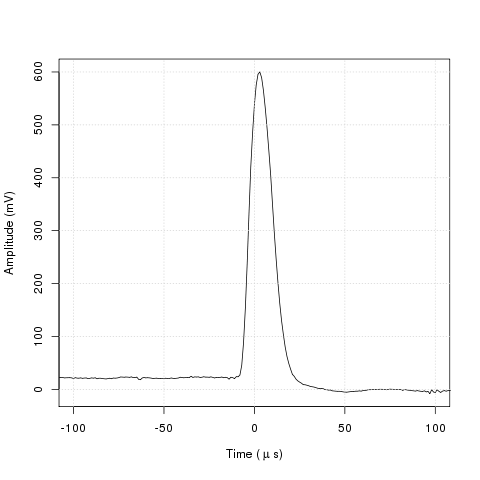} 
\label{fig:demux12}
}
\subfigure[Demultiplexed 486.9~\si{\milli\volt} pulse.]{ 
\includegraphics[width=0.485\linewidth,height=0.306\textheight]{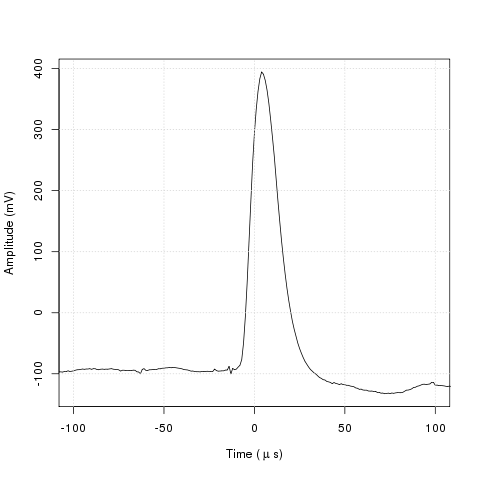} 
\label{fig:demux8}
}\hfil
\subfigure[Demultiplexed 141.1~\si{\milli\volt} pulse.]{ 
\includegraphics[width=0.485\linewidth,height=0.306\textheight]{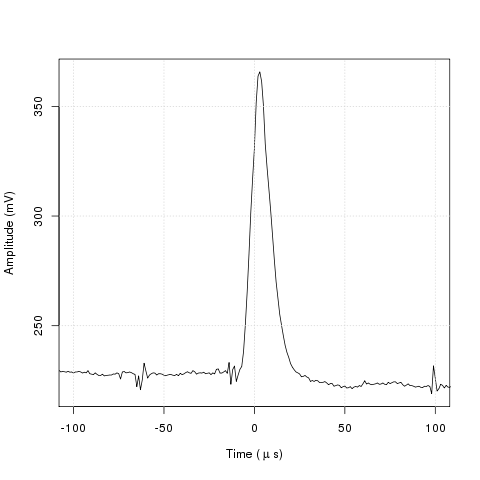} 
\label{fig:demux5}
}
\caption{Reconstructed pulse samples from demultiplexed alpha signals from different detector channels after the low-frequency noise suppression. }
\label{fig:DIIePdemuxpulsesamples}
\end{figure}
To understand the effect of signal multiplexing on the expected intrinsic Bragg curve effect on alpha tracks after these signals were demultiplexed, 1000 alpha event tracks were accumulated. The observed low frequency background noise on each of the signal channels was suppressed by fitting and subtracting 5 harmonics of 5~\si{\kilo\hertz} waveforms on each of the demultiplexed signal channels. This was followed by application of a 30~\si{\micro\second} Savitzky-Golay smoothing filter \cite{Press2007} to further reduce the noise. The Savitzky-Golay filter works by computing the average of polynomial fits on distant data points without distorting the height and width of the original signal. Reconstructed pulse samples from demultiplexed alpha signals are shown for different channels of the detector in Figures \ref{fig:demux4} to \ref{fig:demux5}.

\begin{figure*}[h!]
\centering
\subfigure[Raw signal.]{
\includegraphics[clip, trim=2cm 7cm 3cm 8cm,width=0.49\textwidth,height=0.3\textheight,keepaspectratio]{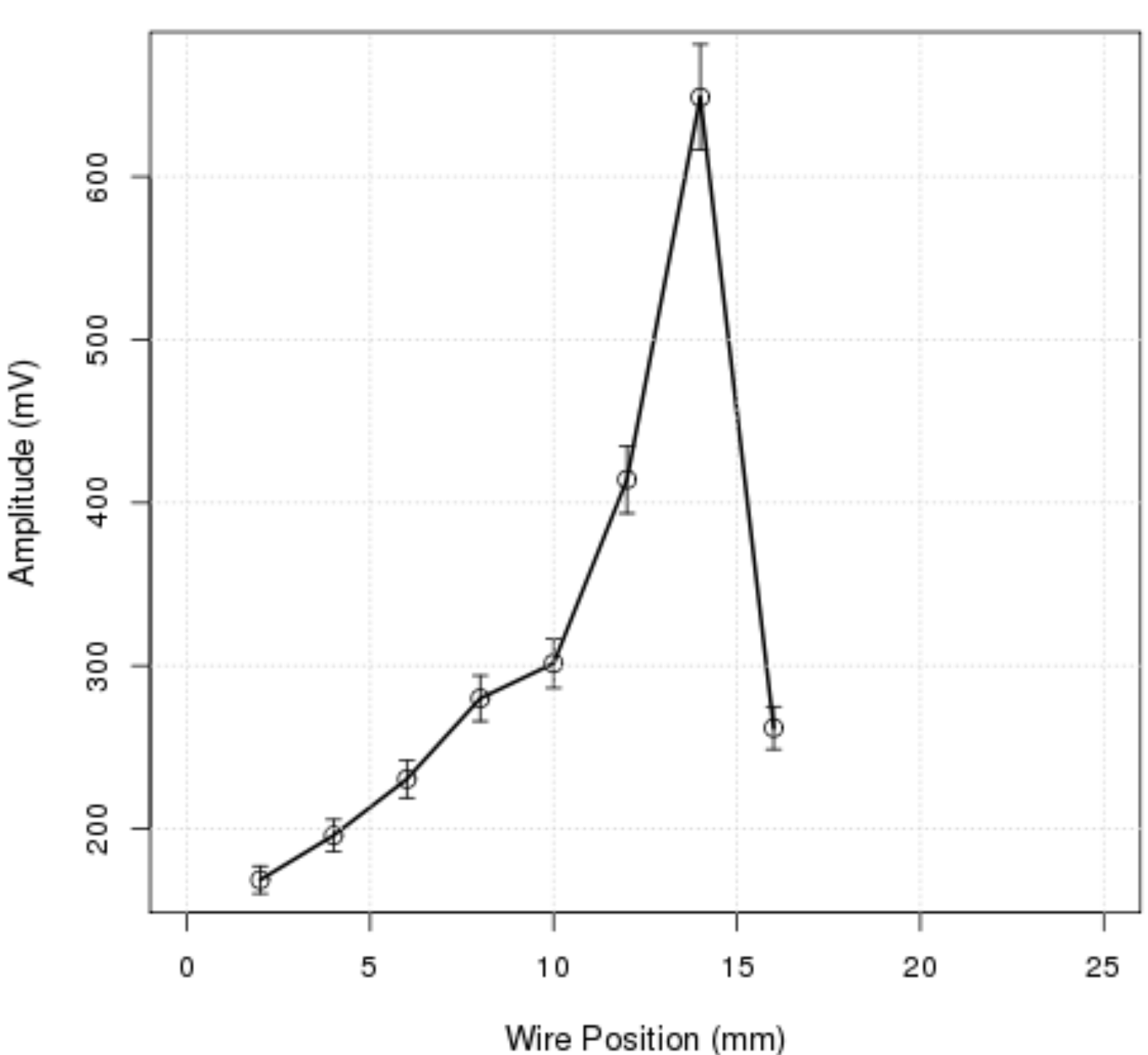}
\label{fig:rawbragg}
}\hfil
\subfigure[Demultiplexed signal.]{
\includegraphics[clip, trim=2cm 7cm 3cm 8cm,width=0.46\textwidth,height=0.3\textheight]{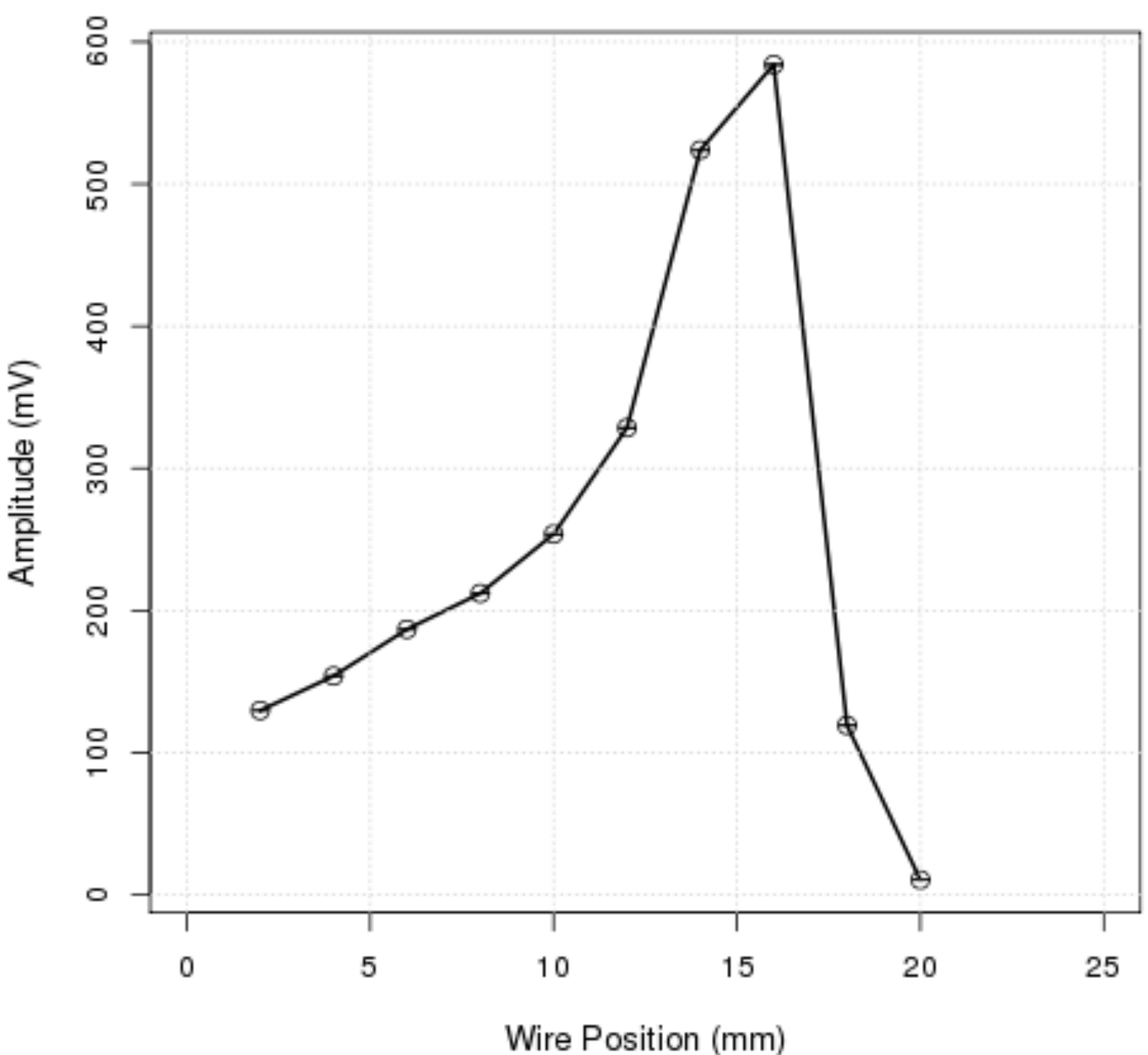}
\label{fig:demuxbragg}
}
\caption{Average event pulse height induced by 5.5~\si{\mega\electronvolt} alpha tracks shown as a function of anode wire separation from the PCB. Low distance values indicate wire closer to the alpha source. Raw signals in 260~\si{\torr} of CF$_4$ are shown in \protect\subref{fig:rawbragg} while \protect\subref{fig:demuxbragg} shows the results from 250~\si{\torr} of CF$_4$ after the signals were multiplexed using the MUX board, demultiplexed and analysed. Quoted errors are 1$\sigma$ statistical uncertainty. }
\label{fig:braggresults}
\end{figure*}

After the noise suppression, the average signal pulse heights recorded on each of the anode wire channels were computed and shown in Figure \ref{fig:demuxbragg}. For comparison, the result from 10 raw detector signals (without any multiplexing and demultiplexing) is shown in Figure \ref{fig:rawbragg}.
The trend of the data in Figure \ref{fig:braggresults} indicate the expected presence of an alpha Bragg peak before and after the multiplexer board was used. The magnitude of the observed Bragg peak is 585$\pm$0.6~\si{\milli\volt} and 650$\pm$32~\si{\milli\volt}, obtained with and without the multiplexer, respectively. Smaller sample size in the measurement without the MUX electronics result in larger uncertainty relative to the MUX result. As expected, each of the Bragg peaks was observed on the anode wire located toward the end of the range of the event track.  It can be seen that only 8 anode data points are shown in Figure \ref{fig:rawbragg}. This is due to the capability of the 16 channeled ADC (8 grid and 8 anode channels) used in this measurement. The $>$10 SNR of the demultiplexed signals is a factor 6 lower than the SNR of raw signals.  Results show that the average magnitude of the observed Bragg peak after the signals were demultiplexed is ~65~\si{\milli\volt} less than the raw signal results. This is mainly due to signal loss in the multiplexer system.  As illustrated in Figure \ref{fig:timingdiagram}, data points that are in-coincidence with the channel or chip switching times are not recorded to disk and may not be recovered in analysis of demultiplexed data set. These dead times of the MUX system and efficiency (>90\si{\percent}) of the noise suppression can contribute to signal loss in the demultiplexed data.  It can be seen that the anode wire (counting from the location closer to the source) on which the Bragg peak was observed is different in the two measurements due to the respective gas pressures used in the two measurements. The energy resolution of the detector was found to be $\sim$13\si{\percent} worse in the demultiplexed signal case relative to the raw signals. Signal losses from the dead time of the DAQ and efficiency of the noise suppression can contribute to this observed difference.  Future work will focus on understanding these effects.

\section{Conclusion}
A one-plane MWPC-based time projection chamber detector was designed, built and used to test the feasibility of a new 20:1 analogue signal multiplexer as a possible readout for a future massive directional dark matter detectors. The 20:1 multiplexer was built using expanded LMH6574 chips from Texas Instruments. Signal multiplexing is motivated and can be a possible means to reduce the cost of signal readouts in massive TPC detectors without compromising the detector sensitivity to event x-y position. Results obtained from this multiplexer system are encouraging and it has demonstrated that ionization charge distribution along alpha tracks can be reconstructed from demultiplexed signals. The precision of the detector energy resolution was found to broaden by $\sim$13\si{\percent} when the multiplexer system was used. Low-frequency harmonic noise, glitches and multiplexer chip switch delays are major factors to considered in future designs. 

\acknowledgments
We are grateful for support from the STFC through ST/P00573X/1 and ST/K001337/1 grants.


\begin{thebibliography}{99}
\bibitem{Mayet2016}
F. Mayet et al.,
\emph{{A review of the discovery reach of directional Dark Matter detection}},
\emph{Phys. Rept.} {\bf 627} (2016) pp 1-49.

\bibitem{Ahlen2010}
S. Ahlen et al.,
\emph{{The Case for a Directional Dark Matter Detector and the Status of Current Experimental Efforts}},
\emph{Int. J. Mod. Phys. A} {\bf 25} 1 (2010) pp 1-51.
	
\bibitem{Ezeribe2017}
A. C. Ezeribe et al.,
\emph{{Demonstration of radon removal from SF$_6$ using molecular sieves}},
\emph{JINST} {\bf 12} 09 (2017) P09025.

\bibitem{Burns2017}
J. Burns et al.,
\emph{{Characterisation of large area THGEMs and experimental measurement of the Townsend coefficients for CF$_4$}},
\emph{JINST} {\bf 12} 10 (2017) T10006.	

\bibitem{Grothaus2014}
P. Grothaus, M. Fairbairn and J. Monroe,
\emph{{Directional dark matter detection beyond the neutrino bound}},
\emph{Phys. Rev. D} {\bf 90} 5 (2014) pp 055018.

\bibitem{Ohare2015}
C. A. J. O'Hare et al.,
\emph{{Readout strategies for directional dark matter detection beyond the neutrino background}},
\emph{Phys. Rev. D} {\bf 92} 6 (2015) pp 063518.

\bibitem{Ohare2016}
C. A. J. O'Hare,
\emph{{Dark matter astrophysical uncertainties and the neutrino floor}},
\emph{Phys. Rev. D} {\bf 94} 6 (2016) pp 063527.

\bibitem{Cushman2013}
Cushman et al.,
\emph{{Working Group Report: WIMP Dark Matter Direct Detection}},
\emph{in Proceedings, 2013 Community Summer Study on the Future of U.S. Particle Physics: Snowmass on the Mississippi (CSS2013): Minneapolis, MN, U.S.A., 29 July - 6 August 2013}, arxiv:1310.8327.

\bibitem{LMH65742014}
Texas Instruments Inc.,
\emph{{Datasheet: LMH6574 4:1 High speed video multiplexer}} (2014).

\bibitem{Horowitz2015}
P. Horowitz and W. Hill,
\emph{{The art of electronics}},
\emph{Cambridge university press}, 3$^{\text{rd}}$ edition, ISBN:~978-0-521-80926-9 (2015).

\bibitem{Stalling2007}
W. Stallings,
\emph{{Data and computer communications}},
\emph{Pearson Prentice Hall}, 8$^{\text{th}}$ edition, ISBN:~0-13-243310-9 (2007).

\bibitem{ni5751-2015}
National Instruments Corporation,
\emph{{Technical Datasheet: NI-5751 Digitizer Adapter Module for FlexRIO}},
\emph{11500 North Mopac Expressway, Austin, Texas 78759, USA} (2015).

\bibitem{niflexrio}
National Instruments Corporation,
\emph{{Technical Datasheet: PXI-7953 NIFlexRIO FPGA Module}},
\emph{11500 North Mopac Expressway, Austin, Texas 78759, USA} (2015).

\bibitem{Cadsoft2014}
CadSoft Computer GmbH,
\emph{{EAGLE: Easily Applicable Graphical Layout Editor, User Language, Version 7.2.0}} (2014), \url{http://eagle.autodesk.com/eagle/documentation}.

\bibitem{Battat2015talk}
J. B. R. Battat, J. Phillips and C. Holman,
\emph{{A 16:1 analog multiplexer: some results and lessons learned}},
\emph{talk presented in DRIFT collaboration meeting held at Occidental College, Los Angeles, U.S.A.} (2015).

\bibitem{Cremat2014shaper}
Cremat,
\emph{{CR-200-4\si{\micro\second} Gaussian Shaping Amplifier:  application guide}} (2014), \url{http://www.cremat.com/CR-200.pdf}. 

\bibitem{Sauli1977}
F. Sauli,
\emph{{Principles of operation of multiwire proportional and drift chambers}},
\emph{CERN-77-09} (1977).

\bibitem{Charpak1979}
G. Charpak and F. Sauli,
\emph{{Multiwire proportional chambers and drift chambers}},
\emph{Nucl. Inst. Meth.} {\bf 162} 1-3 (1979) pp 405 - 428.

\bibitem{Ezeribe2014memo}
A. C. Ezeribe,
\emph{{Construction of DRIFT IIe pro detector with a quad wire winder}},
\emph{an unpublished technical report presented to the DRIFT collaboration} (2014).

\bibitem{Chiba1980}
Y. Chiba and T. Taniguchi,
\emph{{A Simple Semi-Automatic Wire Winding Machine for MWPC}},
\emph{Japanese Journal of Applied Physics} {\bf 19} 6 (1980) pp 1191 - 1192.

\bibitem{Kuze1987}
M. Kuze et al.,
\emph{An Automatic Wire-Winding Machine for MWPC's},
\emph{Japanese Journal of Applied Physics} {\bf 26} 8 (1987) pp 1348 - 1351.

\bibitem{Anderson1992}
W. S. Anderson et al.,
\emph{{Electron attachment, effective ionization coefficient, and electron drift velocity for CF$_4$ gas mixtures}},
\emph{Nuclear Instruments and Methods in Physics Research Section A: Accelerators, Spectrometers, Detectors and Associated Equipment} {\bf 323} 1 (1992) pp 273 - 279.

\bibitem{Cremat2014preamp}
Cremat,
\emph{{CR-111 charge sensitive preamplifier: application guide}} (2014), \url{http://www.cremat.com/CR-111.pdf}. 

\bibitem{Battat2017}
J. B. R. Battat et al.,
\emph{{Low threshold results and limits from the \{DRIFT\} directional dark matter detector}},
\emph{Astroparticle Physics} {\bf 91} (2017) pp 65 - 74.

\bibitem{Veenhof1984}
R. Veenhof,
\emph{{GARFIELD-simulation of gaseous detectors}}, version 9 (2010).

\bibitem{Vavra1993}
J. Va'vra et al.,
\emph{{Measurement of electron drift parameters for helium and CF4-based gases}},
\emph{Nuclear Instruments and Methods in Physics Research Section A: Accelerators, Spectrometers, Detectors and Associated Equipment} {\bf 324} 1 (1993) pp 113 - 126.
	
\bibitem{Lewin1996}
J. D. Lewin and P. Smith,
\emph{{Review of mathematics, numerical factors, and corrections for dark matter experiments based on elastic nuclear recoil}},
\emph{Astroparticle Physics} {\bf 96} 6 (1996) pp 87 - 112.

\bibitem{Nakamura2015}
K. Nakamura et al.,
\emph{{Direction-sensitive dark matter search with gaseous tracking detector NEWAGE-0.3b}},
\emph{Prog. Theor. Exp. Phys.} 043F01 (2015).

\bibitem{Battat2016}
J. B. R. Battat et al.,
\emph{First measurement of nuclear recoil head-tail sense in a fiducialised WIMP dark matter detector},
\emph{JINST} {\bf 11} 10 (2016) P10019.

\bibitem{Battat2017b}
J. B. R. Battat et al.,
\emph{{Measurement of directional range components of nuclear recoil tracks in a fiducialised dark matter detector}},
\emph{JINST} {\bf 12} 10 (2017) P10009.

\bibitem{Ziegler2010}
J. F. Ziegler, M. D. Ziegler and J. P. Biersack,
\emph{{SRIM-The stopping and range of ions in matter (2010)}},
\emph{Nuclear Instruments and Methods in Physics Research Section B: Beam Interactions with Materials and Atoms} {\bf 268} 11 (2010) pp 1818 - 1823.

\bibitem{Cremat2014shaperboard}
Cremat,
\emph{{CR-160-R7 Gaussian shaping amplifier evaluation board: application guide }} (2014), \url{http://www.cremat.com/CR-160-R7.pdf}. 

\bibitem{Press2007}
W. H. Press et al.,
\emph{{Numerical Recipes: The art of scientific computing}},
\emph{Cambridge University Press}, 3$^{\text{rd}}$ edition, ISBN:~13 978-0-511-33555-6 (2007).

\end{thebibliography}
\end{document}